\documentclass[conference]{IEEEtran}
\IEEEoverridecommandlockouts
\usepackage{cite}
\usepackage{amsmath,amssymb,amsfonts}
\usepackage{algorithmic}
\usepackage{graphicx}
\usepackage{textcomp}
\usepackage{xcolor}
\usepackage{fancyhdr}
\usepackage{subcaption}
\usepackage{url}

\makeatletter
\def\ps@IEEEtitlepagestyle{
	\def\@oddhead{\hfill\begin{tabular}{@{}r@{}}\textit{The 30\textsuperscript{th} Annual International Conference of Iranian Society of Mechanical Engineers,} 10 to 12 May, 2022, Tehran, Iran. \\ ISME2022-IC1419\end{tabular}\hfill} 
	\def\@oddfoot{} 
	\def\@evenhead{} 
	\def\@evenfoot{} 
}

\def\ps@plain{\ps@empty 
	\def\@oddhead{} 
	\def\@oddfoot{ 10 to 12 May, 2022\hfill} 
	\def\@evenhead{\hfill} 
	\def\@evenfoot{ 10 to 12 May, 2022\hfill} 
}
\makeatother

\def\BibTeX{{\rm B\kern-.05em{\sc i\kern-.025em b}\kern-.08em
    T\kern-.1667em\lower.7ex\hbox{E}\kern-.125emX}}
\begin{document}
\thispagestyle{IEEEtitlepagestyle} 
\pagestyle{plain} 

\title{Self-Tuning PID Control via a Hybrid Actor-Critic-Based Neural Structure for Quadcopter Control}

\author{
	\IEEEauthorblockN{Iman Sharifi and Aria Alasty\thanks{Iman Sharifi \texttt{\{imansharifi@mech.sharif.edu\}} and Aria Alasty \texttt{\{aalasti@sharif.edu\}}  are with the Department of Mechanical Engineering, Sharif University of Technology, Tehran, Iran.}
	}
}

\maketitle

\begin{abstract}
Proportional–Integral–Derivative (PID) controllers are widely used in industrial and experimental processes due to their simplicity and effectiveness. Traditional offline methods for tuning PID gains, however, are often inadequate for real systems such as quadrotors, which are subject to parameter uncertainties and external disturbances. This paper investigates a self-tuning PID controller for quadrotor attitude and altitude control, based on a reinforcement learning (RL)-driven neural network. An incremental PID structure with both static and dynamic gains is adopted, where only the dynamic gains are adaptively tuned. To achieve this, a model-free hybrid actor–critic neural architecture is employed, enabling both gain tuning and system identification. The proposed method operates online, is computationally efficient, and effectively handles disturbances. Simulation results demonstrate robustness against mass uncertainty and wind gusts, showing that the proposed controller outperforms conventional PID controllers with fixed gains.
\end{abstract}

\section{Introduction}

Due to their low cost, simple mechanical structure, vertical take-off and landing capabilities, and high maneuverability, quadcopters are widely applicable in industrial processes such as agriculture, search and rescue, inspection, and surveillance~\cite{kim2019}. However, because of actuator coupling and the presence of external disturbances, quadcopters exhibit highly nonlinear dynamics. Therefore, attitude control plays a crucial role in trajectory tracking and maneuvering. In~\cite{zulu2016}, a variety of control algorithms were evaluated, showing that none could fully meet the requirements, although hybrid methods demonstrated better adaptability and robustness in the presence of disturbances.

The PID controller is extensively used in industrial systems due to its simplicity and ease of implementation. Nevertheless, its accuracy is highly dependent on controller gains and the system model. In high-order nonlinear systems, parameter uncertainties and external disturbances can significantly degrade PID performance. Conventional offline tuning methods are not efficient for such systems~\cite{pan2009}. To achieve acceptable performance in nonlinear settings, online methods such as Adaptive Control, Fuzzy Systems, and Neural Networks (NNs) are preferred~\cite{yang2013,goel2020,hernandez2016,bari2019,park2021}. NNs are capable of solving non-trivial problems efficiently and approximating high-order nonlinear functions~\cite{abiodun2018}. 

Furthermore, Reinforcement Learning (RL) methods, which often incorporate NNs in their structure, have demonstrated remarkable capabilities and, in some cases, even surpass human performance. RL algorithms are semi-supervised learning approaches in which an agent learns through interactions with its environment. RL has considerable utility in self-tuning PID control and can operate without human intervention. For instance, the Q-learning algorithm has been employed for PID gain tuning~\cite{carlucho2017,shi2018}. However, Q-learning cannot handle continuous actions and requires substantial computational resources to achieve high accuracy. Actor-Critic methods, on the other hand, allow agents to generate continuous control signals and, importantly, provide an online and NN-based solution. In~\cite{guan2021}, a Radial Basis Function (RBF) NN was used for actor policy and critic value function approximation, demonstrating that this approach can track complex trajectories. Deep Deterministic Policy Gradient (DDPG) has also been applied to PID tuning~\cite{lawrence2022,yang2021}, but it relies on offline training and demands significant computational power. Moreover, the performance of DDPG-trained models may deteriorate in real systems subject to environmental disturbances. In~\cite{sun2021}, the Asynchronous Advantage Actor-Critic (A3C) method was employed for PID tuning, enabling multi-actor and multi-critic learning. Results showed that this approach improved PID controller performance.

In this paper, a new structure for tuning PID gains is introduced using neural networks (NNs), leveraging recent algorithms capable of performing both self-tuning PID control and system identification. This fast and online method does not require large memory capacity, powerful processors, or offline training. The Adaptive Moment Estimation (ADAM) optimizer is employed to update the network weights using the Backpropagation algorithm~\cite{kingma2014}. ADAM is known for its efficiency in deep networks, speed, and ability to escape shallow local minima. To evaluate the effectiveness of the proposed method, we examine its performance under conditions of mass uncertainty and wind gust disturbances.

The remainder of this paper is organized as follows. Section~II presents the dynamical modeling of the quadrotor and the PID control method. Section~III describes the design of the hybrid neural structure for online PID gain tuning, followed by optimization. Section~IV provides comparative numerical simulations to demonstrate the effectiveness of the proposed controller. Finally, Section~V summarizes the conclusions and highlights the contributions of this work.

\section{Dynamic Modeling and PID Control Method}

\begin{figure}[htbp]
	\centerline{\includegraphics{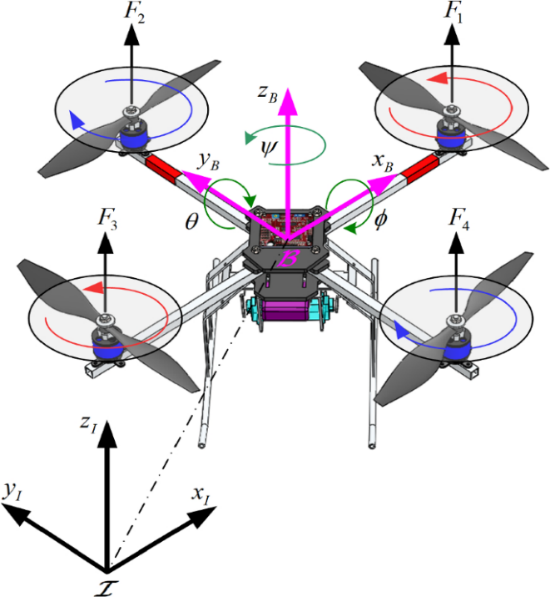}}
	\caption{Schematic of the quadcopter.}
	\label{quad}
\end{figure}

The quadcopter (as shown in Fig.~\ref{quad}) is an under-actuated system with four control inputs ($u_1, u_2, u_3, u_4$) and six degrees of freedom (DOFs), including position ($x, y, z$) and attitude ($\phi, \theta, \psi$). Due to the nonlinearities of the quadcopter system and the complexity of environmental conditions, it is nearly impossible to model this robot with complete accuracy. In such cases, system identification methods such as neural networks (NNs) can efficiently estimate the system states. Therefore, our control method does not require an exact model and relies only on instantaneous inputs and outputs of the system.

In this research, a simplified mathematical model~\cite{tripathi2015} with known parameters is used solely to simulate the real system in the absence of noise. In practice, we assume that the system parameters are unknown and estimate the states using the corresponding control inputs and recent states. The governing equations are given in Eq.~\ref{quad eqs}. In this formulation, $x, y, z$ denote the positions of the center of gravity relative to the inertial reference coordinates ($x_I, y_I, z_I$), and $\phi, \theta, \psi$ represent the rotational angles about the body axes ($x_B, y_B, z_B$):
\begin{equation}
	\label{quad eqs}
	\begin{aligned}
		\ddot{\phi} &= \dot{\theta}\dot{\psi}\frac{J_y - J_z}{J_x} + \frac{l}{J_x}u_2, \\	
		\ddot{\theta} &= \dot{\phi}\dot{\psi}\frac{J_z - J_x}{J_y} + \frac{l}{J_y}u_3, \\
		\ddot{\psi} &= \dot{\phi}\dot{\theta}\frac{J_x - J_y}{J_z} + \frac{1}{J_z}u_4, \\
		\ddot{z} &= \frac{u_1}{m}\cos\phi\cos\theta - g, \\
		\ddot{x} &= \frac{u_1}{m}\big(\cos\phi\sin\theta\cos\psi + \sin\phi\sin\psi\big), \\
		\ddot{y} &= \frac{u_1}{m}\big(\cos\phi\sin\theta\sin\psi - \sin\phi\cos\psi\big),
	\end{aligned}
\end{equation}		
where $m$ is the total mass of the robot, $g$ is gravitational acceleration, $l$ is the quadcopter arm length, and $J_x, J_y, J_z$ are the moments of inertia about the body axes, respectively.
Control inputs (Eq.~\ref{control inputs}) are expressed as combinations of the squared motor angular velocities ($\Omega_1, \Omega_2, \Omega_3, \Omega_4$):  
\begin{equation}
	\label{control inputs}
	\begin{aligned}
		u_1 &= b\big(\Omega_1^2 + \Omega_2^2 + \Omega_3^2 + \Omega_4^2\big), \quad \quad\\
		u_2 &= b\big(\Omega_4^2 - \Omega_2^2\big), \\
		u_3 &= b\big(\Omega_3^2 - \Omega_1^2\big), \\
		u_4 &= d\big(\Omega_4^2 + \Omega_2^2 - \Omega_1^2 + \Omega_3^2\big),
	\end{aligned}
\end{equation}
where $b$ and $d$ denote the thrust and torque coefficients, respectively.  

To determine the control inputs in an online manner, a PID control algorithm is employed. First, the static PID gains are selected using either trial and error or the Ziegler–Nichols method, ensuring that the system achieves acceptable stability. These static gains remain fixed throughout each mission. The proposed method then tunes the dynamic gains, which are added to the corresponding static gains. Thus, each control input ($u_1, u_2, u_3, u_4$) can be expressed as the summation of static and dynamic components:  
\begin{equation}
	u(t) = u_{sg}(t) + u_{dg}(t),
\end{equation}
where
\begin{equation}
	\begin{aligned}
		u_{sg}(t) &= K_p^s e(t) + K_i^s \int_{0}^{t} e(\tau)\, d\tau + K_d^s \dot{e}(t), \\
		u_{dg}(t) &= K_p^d e(t) + K_i^d \int_{0}^{t} e(\tau)\, d\tau + K_d^d \dot{e}(t).
	\end{aligned}
\end{equation}

Here, $K_p^s, K_i^s, K_d^s$ are constant gains determined in advance, while $K_p^d, K_i^d, K_d^d$ are dynamic gains tuned online using a neural network based on the actor–critic method. This approach will be elaborated upon in the subsequent sections.

\section{Network Structure}

The proposed method consists of two main components: (i) a self-tuning PID controller, and (ii) a system identification module based on neural networks (NNs). In the first part, an NN is designed to tune the dynamic PID gains. The control input is then computed by externally injecting PID errors into the network. At each step, the obtained control input, along with recent system outputs, is fed into the system identification network. This network estimates the new system output using an Actor–Critic structure. Specifically, the identification module is composed of two networks: an Actor network, which attempts to identify the actual system output, and a Critic network, which computes the value function of the inputs (environment states) and evaluates the quality of the Actor’s action in the current state.

\begin{figure}[htbp]
	\centerline{\includegraphics[width=\linewidth]{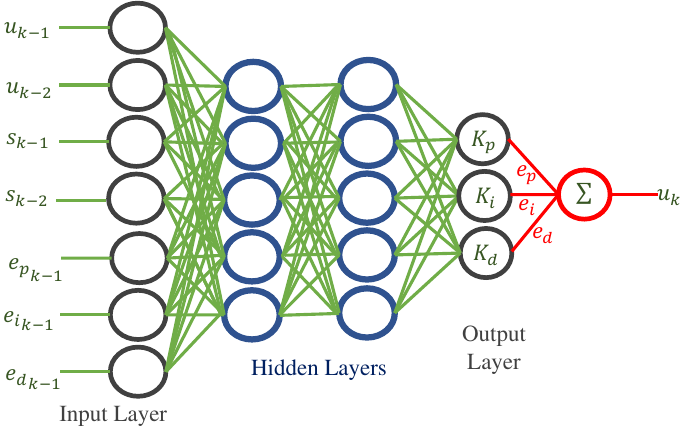}}
	\caption{Self-tuning PID neural network. Inputs are the current control inputs, current states, and PID errors. Hidden layers use the sigmoid activation function, while the output layer employs the $\tanh$ activation function.}
	\label{stpid}
\end{figure}

As illustrated in Fig.~\ref{stpid}, the sigmoid activation function is applied in the hidden layers, whereas the $\tanh$ activation function is employed in the PID gains layer. Finally, the dynamic PID gains ($K_n^d$) are computed as:
\begin{equation}
	\begin{aligned}
		K_n^d(k) = f_n \big(&u(k-1), u(k-2), s(k-1), s(k-2), \\
		               &e_p(k-1), e_i(k-1), e_d(k-1)\big),
	\end{aligned}
	\label{PIDgains}
\end{equation}
where $n \in \{p, i, d\}$, and $f_n(\cdot)$ is a nonlinear function parameterized by numerous weights and biases, which are initialized within a small bound near zero.
In the identification network, the Actor network has two outputs: a mean ($\mu$) and a variance ($\sigma$). These outputs are used to parameterize a normal distribution $\mathcal{N}(\mu,\sigma^2)$, from which a sample is drawn randomly.  

\begin{figure}[htbp]
	\centerline{\includegraphics[width=\linewidth]{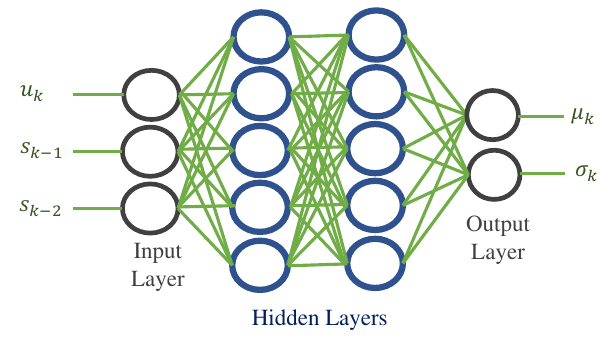}}
	\caption{Actor network. Inputs are the current control input and current states. Hidden layers use the sigmoid activation function.}
	\label{actor}
\end{figure}

Here, $s_m$ represents the estimated state of the quadcopter (attitude and altitude). This sampled value constitutes the final output of the Actor network, which is expected to track the actual system output.  

The Critic network estimates the value function ($v$) using the system states (control input and recent outputs). In doing so, it provides feedback to the Actor, enabling it to improve its performance.  

\begin{figure}[htbp]
	\centerline{\includegraphics[width=\linewidth]{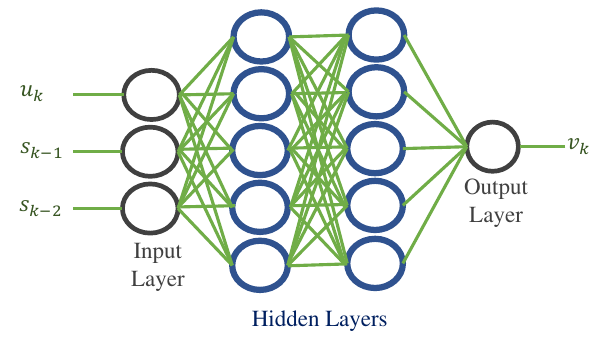}}
	\caption{Critic network. Inputs are the current control input and system states. Hidden layers use the sigmoid activation function, and the output is the value function.}
	\label{critic}
\end{figure}

\begin{figure*}[htbp]
	\centerline{\includegraphics[width=\textwidth]{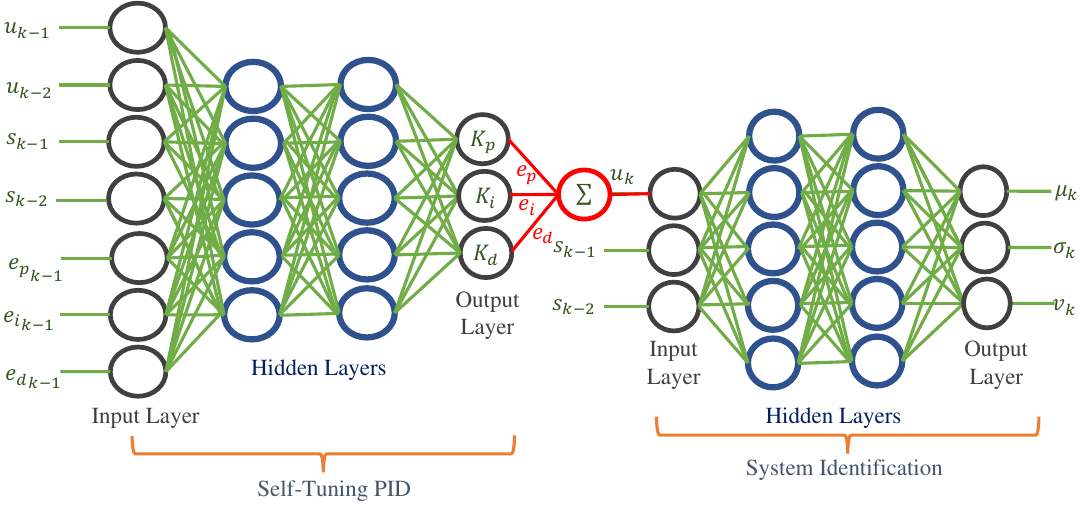}}
	\caption{Overall architecture of the proposed network. The left sub-network tunes PID gains for each quadcopter state and computes the corresponding control input. The right sub-network is the Actor–Critic module, which estimates each state.}
	\label{general}
\end{figure*}

Finally, the outputs of the system identification network are obtained using the following equations:
\begin{align}
	\mu(k) &= f_\mu\big(u(k), s(k-1), s(k-2)\big), \\
	\sigma(k) &= f_\sigma\big(u(k), s(k-1), s(k-2)\big), \\
	v(k) &= f_v\big(u(k), s(k-1), s(k-2)\big),
\end{align}
where $f_\mu(\cdot)$ and $f_\sigma(\cdot)$ denote the Actor functions, and $f_v(\cdot)$ denotes the Critic function.  

Having designed the self-tuning PID, Actor, and Critic networks individually, we now connect them to operate jointly in order to achieve the final objective of self-tuning with system identification. Specifically, the self-tuning PID network is connected in series with the system identification network (Fig.~\ref{general}). In this configuration, the output of the self-tuning PID network serves as the input to the identification network, forming a unified architecture. The inputs to this combined network are the control inputs, current states, and PID errors, while the outputs are the estimated system states and the value function of the inputs. Consequently, no explicit system model is required, making the proposed method entirely model-free.

\section{Optimization}

After designing the network structure and initializing the weights and biases, the network parameters must be adjusted using an optimizer. The Actor’s objective is output estimation, i.e., minimizing the estimation error $(s_m - s)$, while the Critic aims to minimize the Temporal Difference (TD) error $\delta_{TD}$~\cite{sutton2018}. The TD error is computed as:
\begin{equation}
	\delta_{TD} = R_{k+1} + \gamma v_{k+1} - v_{k},
\end{equation}
where $\gamma$ is the discount factor, $R_{k+1}$ is the reward function, and $v_k$ and $v_{k+1}$ are the value functions at steps $k$ and $k+1$, respectively. The reward function is defined in quadratic form so that smaller absolute error, error rate, and control effort result in higher rewards:  
\begin{equation}
	R_{k+1} = -r_1 (s_m - s)^2 - r_2 (\dot{s}_m - \dot{s})^2 - r_3 u^2,
\end{equation}
where $r_1, r_2, r_3$ are weighting coefficients for the error, error rate, and control input, respectively, and $u$ is the control signal.  

The loss functions for the Actor ($\mathcal{L}_a$) and Critic ($\mathcal{L}_c$) are defined as follows:  
\begin{equation}
	\label{loss eqs}
	\begin{aligned}
		\mathcal{L}_a &= w_1 (s_m - s)^2 \big(\eta + |\delta_{TD}|\big) + w_2 \sqrt{2 \pi e \sigma^2}, \\
		\mathcal{L}_c &= w_3 \delta_{TD}^2,
	\end{aligned}
\end{equation}

where $w_1, w_2, w_3$ are constant weights that determine the relative importance of each term. The parameter $\eta$ is a small constant introduced to prevent the Actor’s loss from reaching zero when $\delta_{TD}$ approaches zero. This ensures that the Actor continues to explore the environment until convergence to the optimal solution.  

\begin{figure*}[t]
\centering
\begin{subfigure}{0.48\linewidth}
    \centering
    \includegraphics[width=\linewidth]{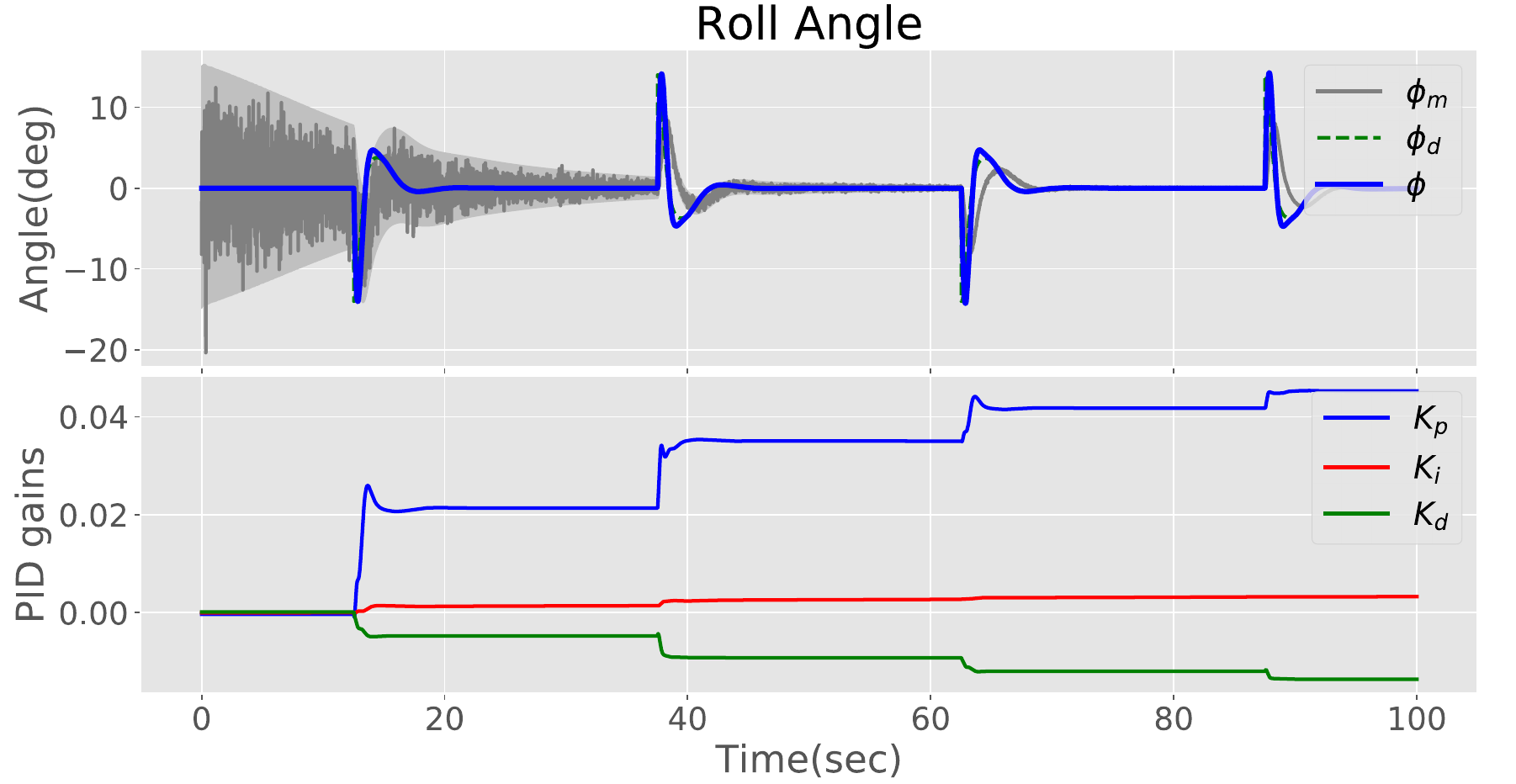}
    \caption{Roll ($\phi$) angle control.}
    \label{roll0}
\end{subfigure}
\hfill
\begin{subfigure}{0.48\linewidth}
    \centering
    \includegraphics[width=\linewidth]{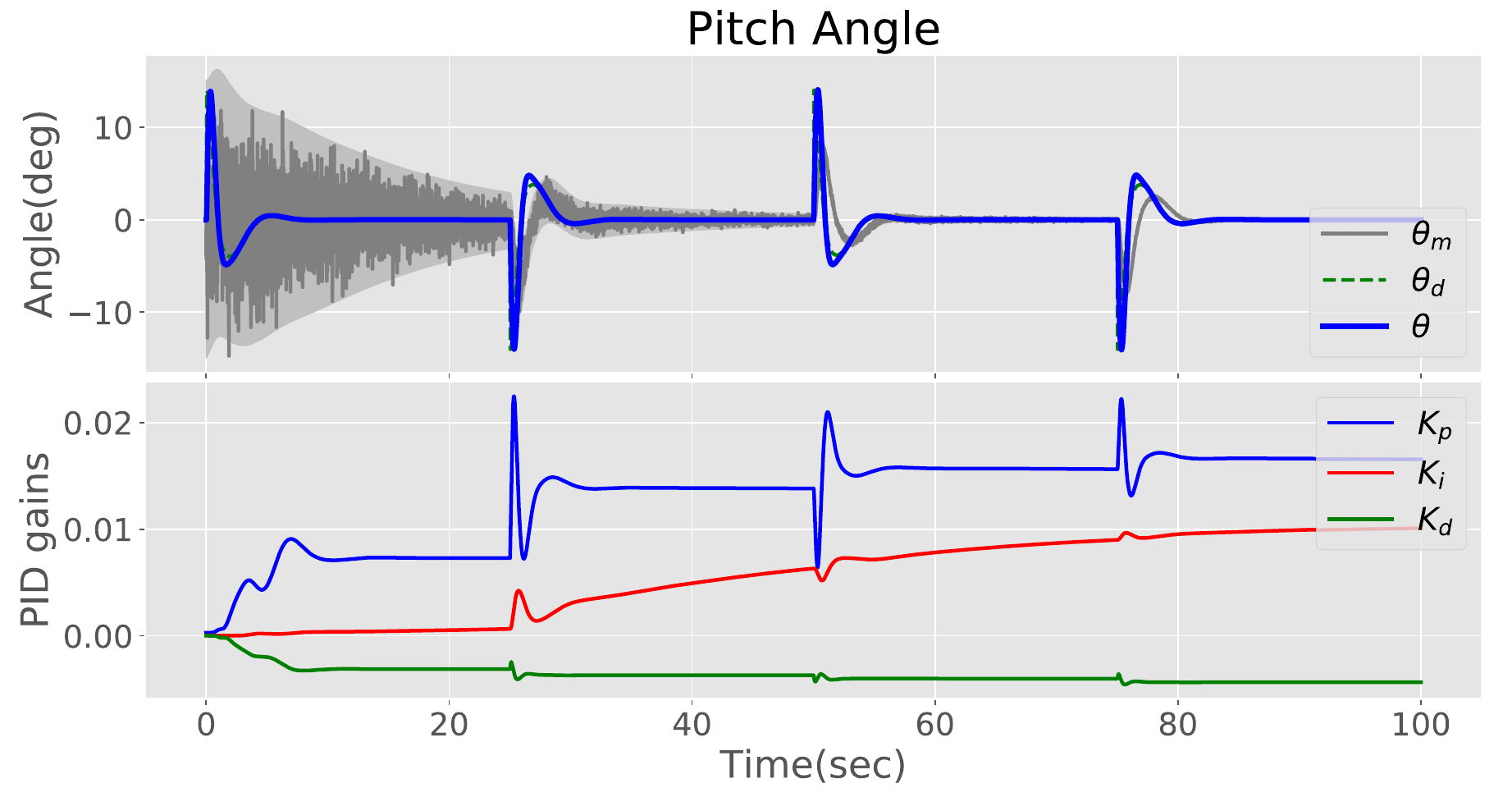}
    \caption{Pitch ($\theta$) angle control.}
    \label{pitch0}
\end{subfigure}

\vspace{0.3cm}

\begin{subfigure}{0.48\linewidth}
\centering
\includegraphics[width=\linewidth]{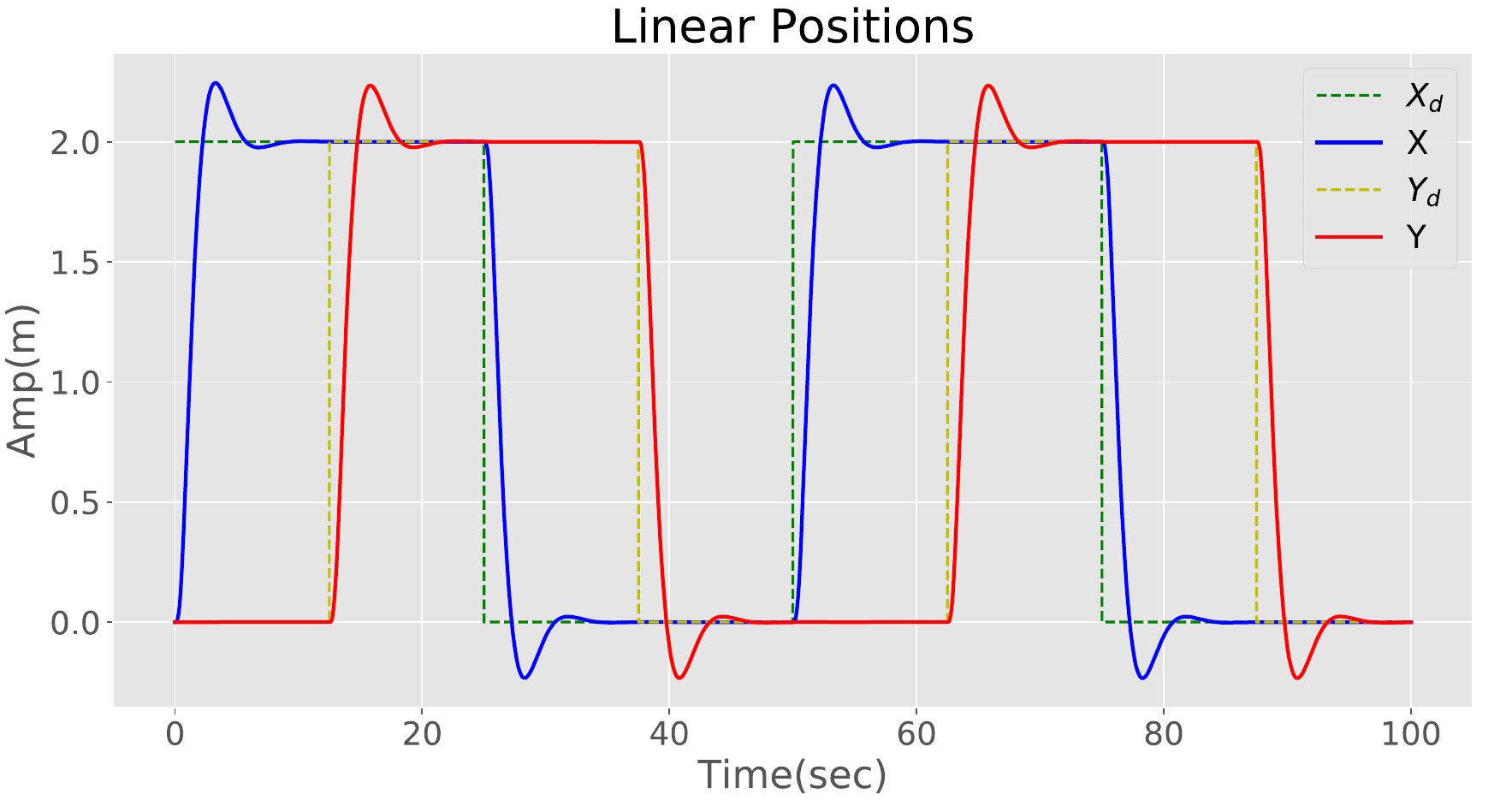}
\caption{X–Y position tracking along the squared path.}
\label{xy0}
\end{subfigure}
\hfill
\begin{subfigure}{0.48\linewidth}
    \centering
    \includegraphics[width=\linewidth]{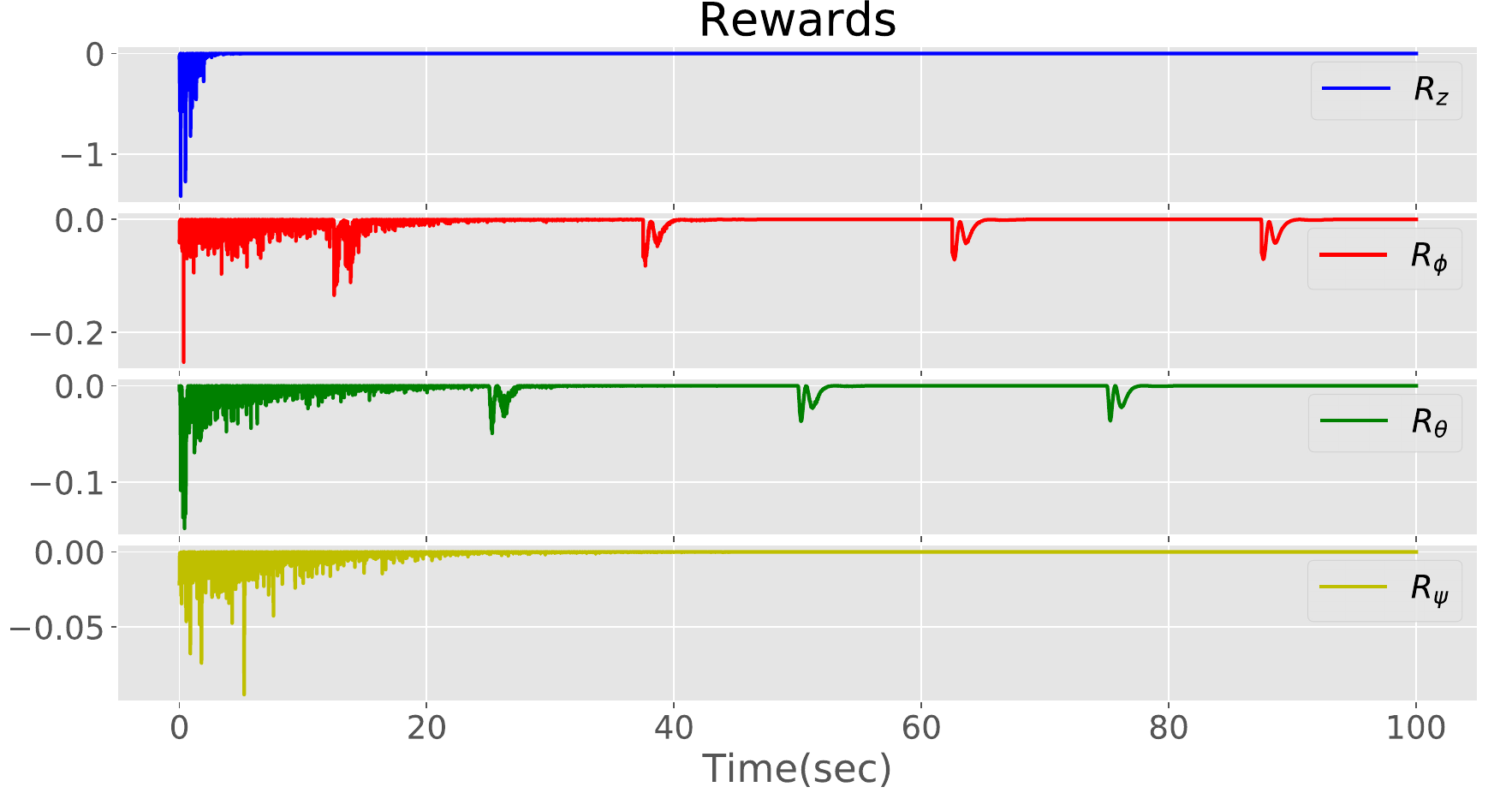}
    \caption{Rewards during training.}
    \label{rewards}
\end{subfigure}

\caption{Performance of the proposed controller in the squared path scenario: (a) Roll angle tracking, (b) Pitch angle tracking, (c) X-Y positions, and (d) agent rewards.}
\label{squarepath_results}
\end{figure*}
The Critic computes the $\delta_{TD}$ signal and provides it to the Actor as feedback. If $\delta_{TD}$ is large, the Actor’s loss will also be large, encouraging the Actor to select improved actions. Conversely, when $\delta_{TD}$ is close to zero, no significant change in action is required, leading to the convergence of the Actor’s policy to the optimal solution.  

The complete network, which integrates both the self-tuning PID module and the system identification network, is shown in Fig.~\ref{general}.
A general loss function ($\mathcal{L}_t$) is required to optimize the network, which is obtained by combining the Actor and Critic loss functions:  
\begin{equation}
	\mathcal{L}_t = \mathcal{L}_a + \mathcal{L}_c.
\end{equation}

To update the network parameters, the ADAM optimizer is employed, as it is both efficient and reliable for deep neural networks. The update equations of ADAM are given in Eq.~\ref{adam eqs}:  
\begin{equation}
	\label{adam eqs}
	\begin{aligned}
		m_t &= \beta_1 m_t + (1 - \beta_1) g_t, \\
		v_t &= \beta_2 v_t + (1 - \beta_2) g_t^{2}, \\
		\hat{m}_t &= \frac{m_t}{1 - \beta_1}, \quad \hat{v}_t = \frac{v_t}{1 - \beta_2}, \\
		\theta_{t+1} &= \theta_t - \frac{\alpha}{\sqrt{\hat{v}_t + \epsilon}} \hat{m}_t,
	\end{aligned}
\end{equation}
where $g_t = \frac{\partial \mathcal{L}_t}{\partial \theta_t}$ is the gradient of the total loss with respect to the parameters. Let $\theta_{st}$ and $\theta_{si}$ denote the weights of the self-tuning and system identification networks, respectively. The gradients of the total loss function with respect to these parameters are computed as follows:  
\begin{align}
	g_{st} &= \left( 
	    \frac{\partial \mathcal{L}_a}{\partial s_m} \frac{\partial s_m}{\partial u} 
	    + \frac{\partial \mathcal{L}_a}{\partial \sigma} \frac{\partial \sigma}{\partial u} 
	    + \frac{\partial \mathcal{L}_c}{\partial v} \frac{\partial v}{\partial u} 
	\right) \frac{\partial u}{\partial \theta_{st}}, \\
	g_{si} &= 
	    \frac{\partial \mathcal{L}_a}{\partial s_m} \frac{\partial s_m}{\partial \theta_{si}} 
	    + \frac{\partial \mathcal{L}_a}{\partial \sigma} \frac{\partial \sigma}{\partial \theta_{si}} 
	    + \frac{\partial \mathcal{L}_c}{\partial v} \frac{\partial v}{\partial \theta_{si}}.
\end{align}

In particular,
\begin{equation}
	\frac{\partial u}{\partial \theta_{st}} 
	= e_p \frac{\partial K_p^d}{\partial \theta_{st}} 
	+ e_i \frac{\partial K_i^d}{\partial \theta_{st}} 
	+ e_d \frac{\partial K_d^d}{\partial \theta_{st}},
\end{equation}
where $e_p, e_i, e_d$ are externally injected error signals that are not updated by the optimizer.  

The proposed structure is directly applicable to Single-Input–Single-Output (SISO) systems, whereas the quadcopter is a Multi-Input–Multi-Output (MIMO) system. Fortunately, by assuming that the states remain close to the equilibrium point, the system can be decoupled into four independent SISO subsystems: $\phi, \theta, \psi$, and $z$. Among these, $\phi$ and $\theta$ are underactuated subsystems, and the translational states $x$ and $y$ are fully dependent on them.  

\section{Results}
After designing the network structure and setting up the optimizer, we simulated the network using Python and the PyTorch library. Additionally, V-REP CoppeliaSim\footnote{\url{https://www.coppeliarobotics.com/simulator}} was employed as a realistic simulation environment. In this framework, the model of the quadcopter is assumed to be completely unknown, and only the system inputs and outputs are continuously available. Several scenarios were considered to verify and challenge the proposed method. 

\begin{figure*}[t]
\centering
\begin{subfigure}{0.48\linewidth}
    \centering
    \includegraphics[height=5cm]{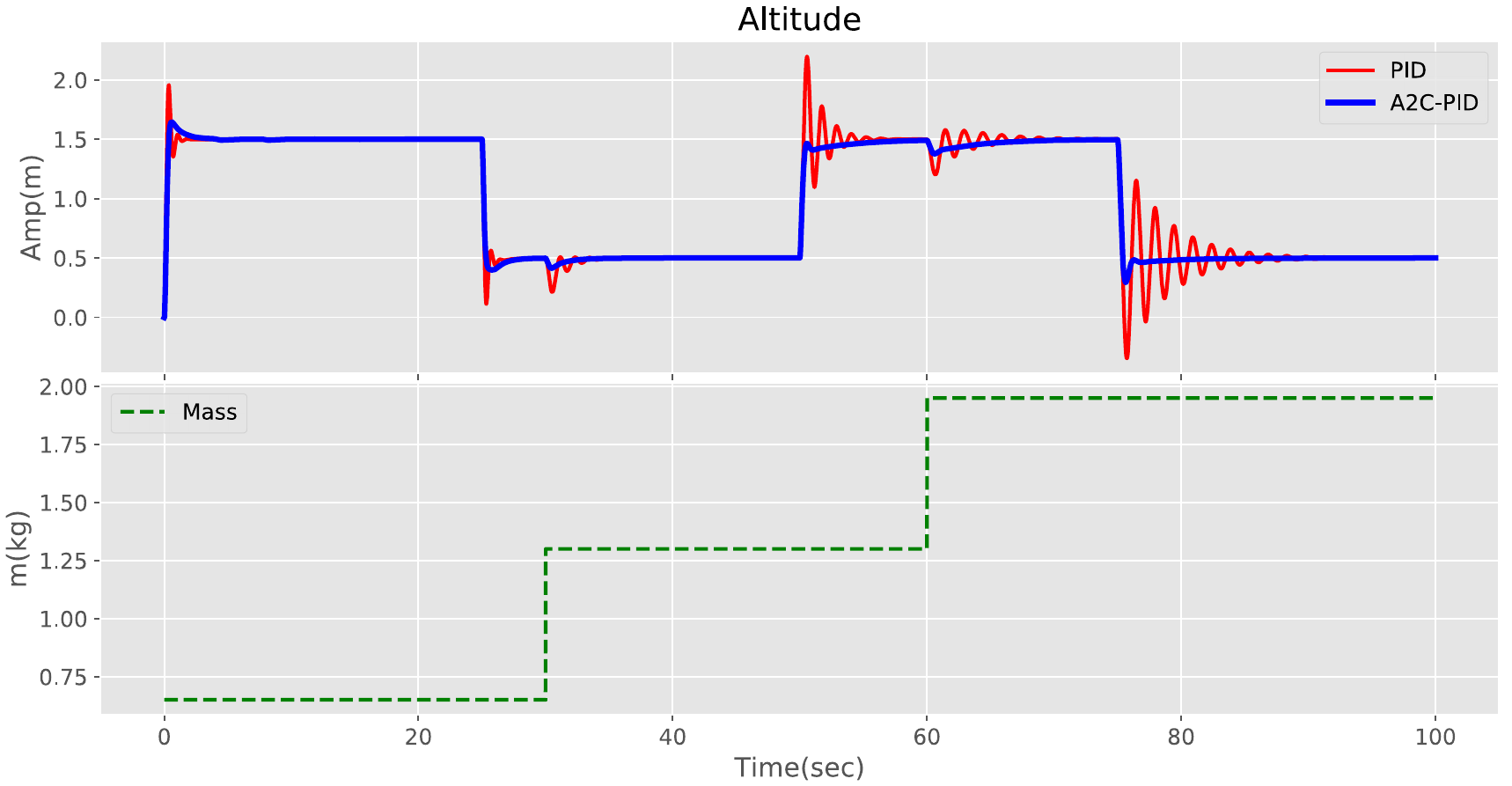}
    \caption{Altitude control under mass variation.}
    \label{mass}
\end{subfigure}
\hfill
\begin{subfigure}{0.48\linewidth}
    \centering
    \includegraphics[height=5cm]{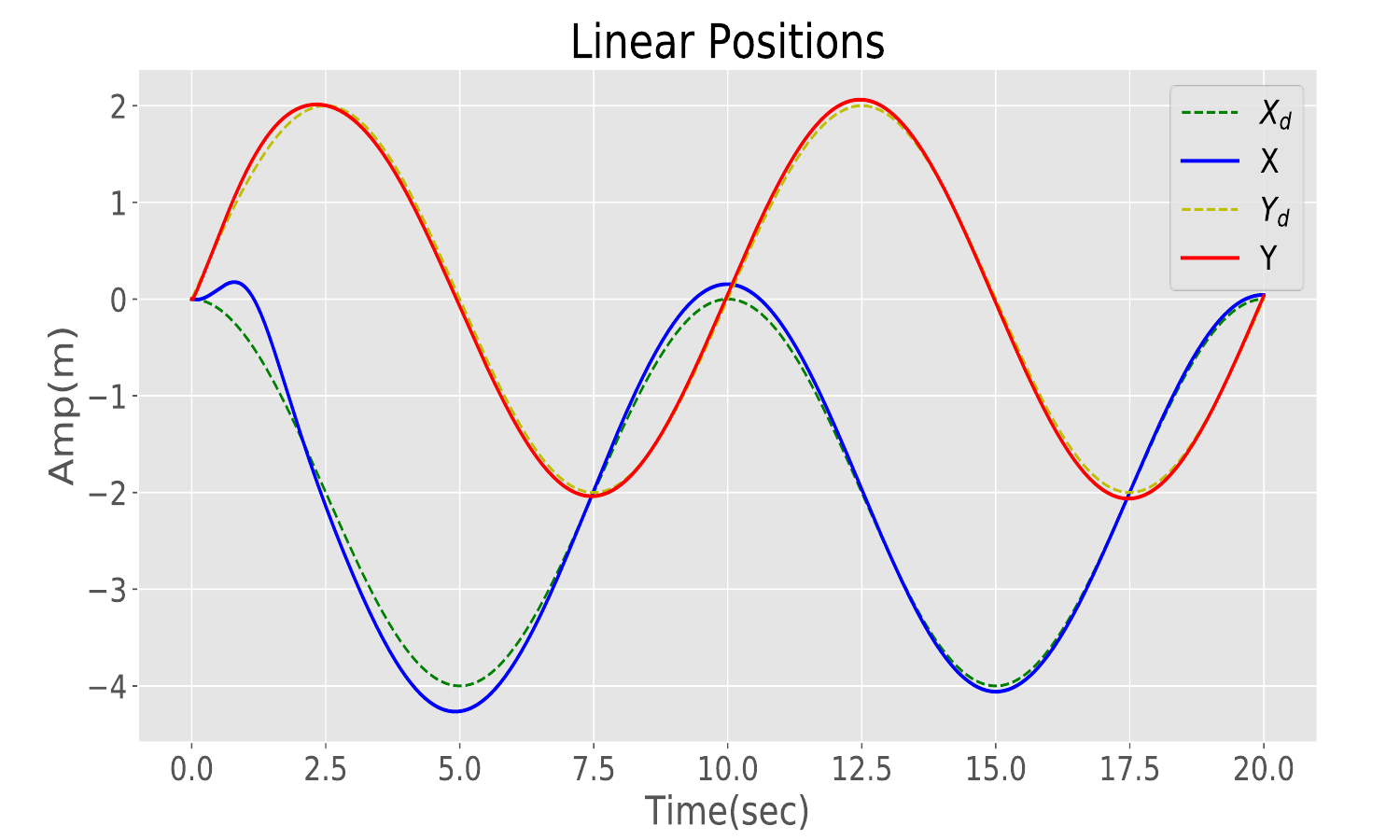}
    \caption{X–Y position tracking along a helical path.}
    \label{xy1}
\end{subfigure}
\caption{Performance of the proposed controller: (a) mass variation, (b) helical path.}
\end{figure*}

\subsection*{A. Squared Path at Constant Height}

As a first test, a squared path at constant altitude was chosen as the tracking scenario. Fig.~\ref{roll0} and Fig.~\ref{pitch0} show that, at the beginning of the mission, the variance ($\sigma$) of the gain selection is high, causing the agent to explore different actions to improve identification. Consequently, the rates of change of the gains are initially large but gradually decrease and converge to constant values. These results demonstrate that the proposed method can effectively control the attitude of the quadcopter in a simple trajectory-tracking task.

Fig.~\ref{rewards} presents the rewards gained by the agents during training. Both metrics increase over time and approach their optimal values, confirming that the network weights were successfully optimized. Since the networks are trained in an online manner, the method is computationally efficient and can be applied to real-world robotic systems.  

By tuning the PID gains in this scenario, the attitude control (Euler angles) successfully tracked the desired references. Once satisfactory attitude control was achieved, position tracking could also be obtained, as illustrated in Fig.~\ref{xy0}.  

\subsection*{B. Mass Uncertainty}

In the next scenario, the total mass of the quadcopter was varied with respect to time (Fig.~\ref{mass}) to introduce parameter uncertainty and evaluate the robustness of the proposed network. The system mass changes significantly after a short period, causing corresponding variations in altitude. This situation requires the controller to adapt to the new conditions. As shown in Fig.~\ref{mass}, the proposed controller successfully compensates for the resulting error.  
Compared to a conventional PID controller, it is evident that the proposed method performs better under mass uncertainty. With additional mass, the performance of the traditional PID controller deteriorates because its gains are designed for a fixed system configuration. In contrast, the proposed method dynamically adjusts its gains, preventing error growth and maintaining stability. Thus, the approach is not only online but also inherently adaptable to system variations.  

As a more complex test, a helical path was designed to further evaluate the control performance. For an underactuated system such as a quadcopter, successful position tracking in multiple directions implies that attitude control—the inner loop of the system—has also been achieved. Fig.~\ref{xy1} illustrates that the robot tracks the helical path effectively, with the tracking error converging toward zero. When position tracking is accurate, attitude control is also ensured.  

\subsection*{C. Disturbance Rejection}

\begin{figure*}[htbp]
\centering
\begin{subfigure}{0.48\linewidth}
    \centering
    \includegraphics[width=\linewidth]{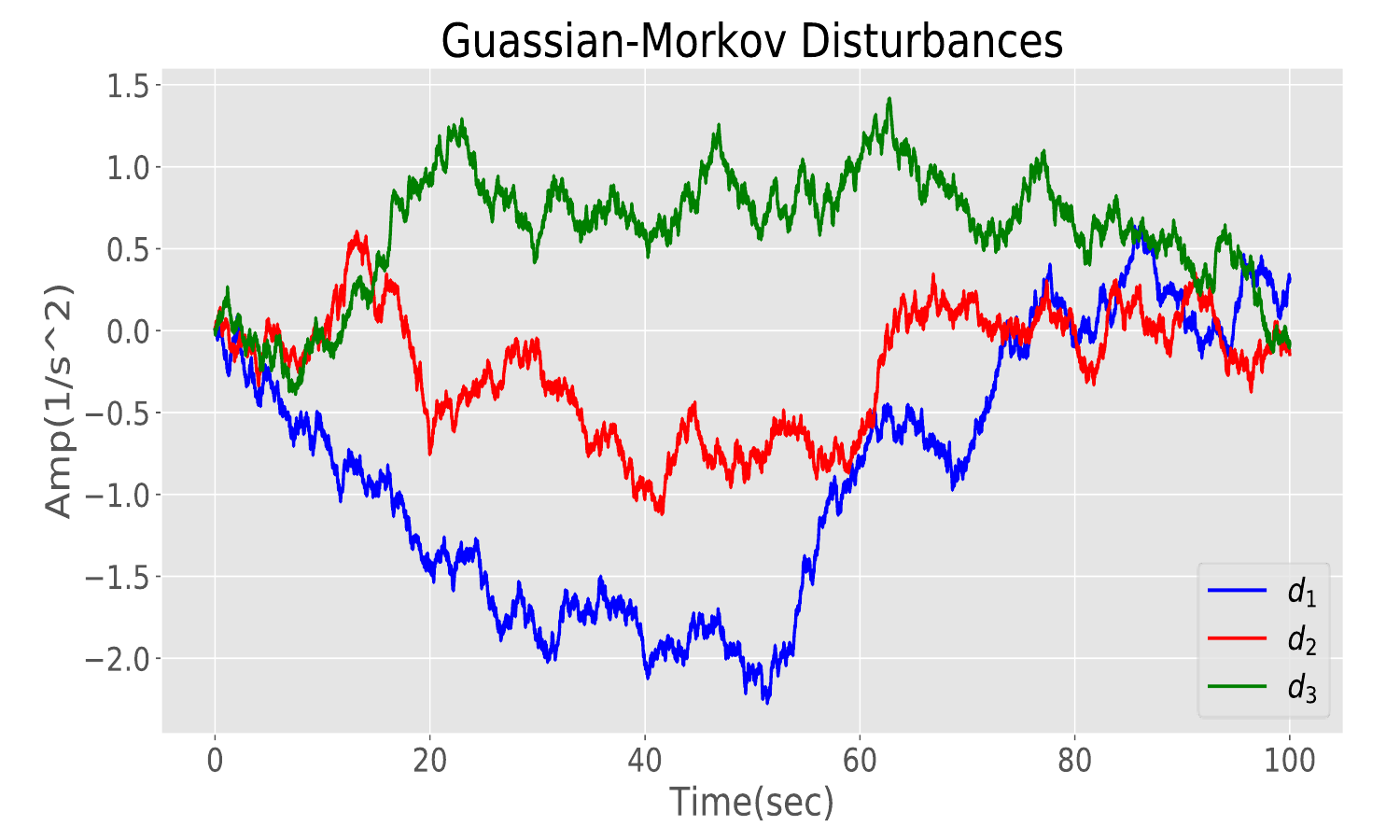}
    \caption{Gaussian–Markov disturbances. $d_1, d_2, d_3$ correspond to $\phi, \theta, \psi$ and are added to these angles, respectively.}
    \label{dist}
\end{subfigure}
\hfill
\begin{subfigure}{0.48\linewidth}
    \centering
    \includegraphics[width=\linewidth]{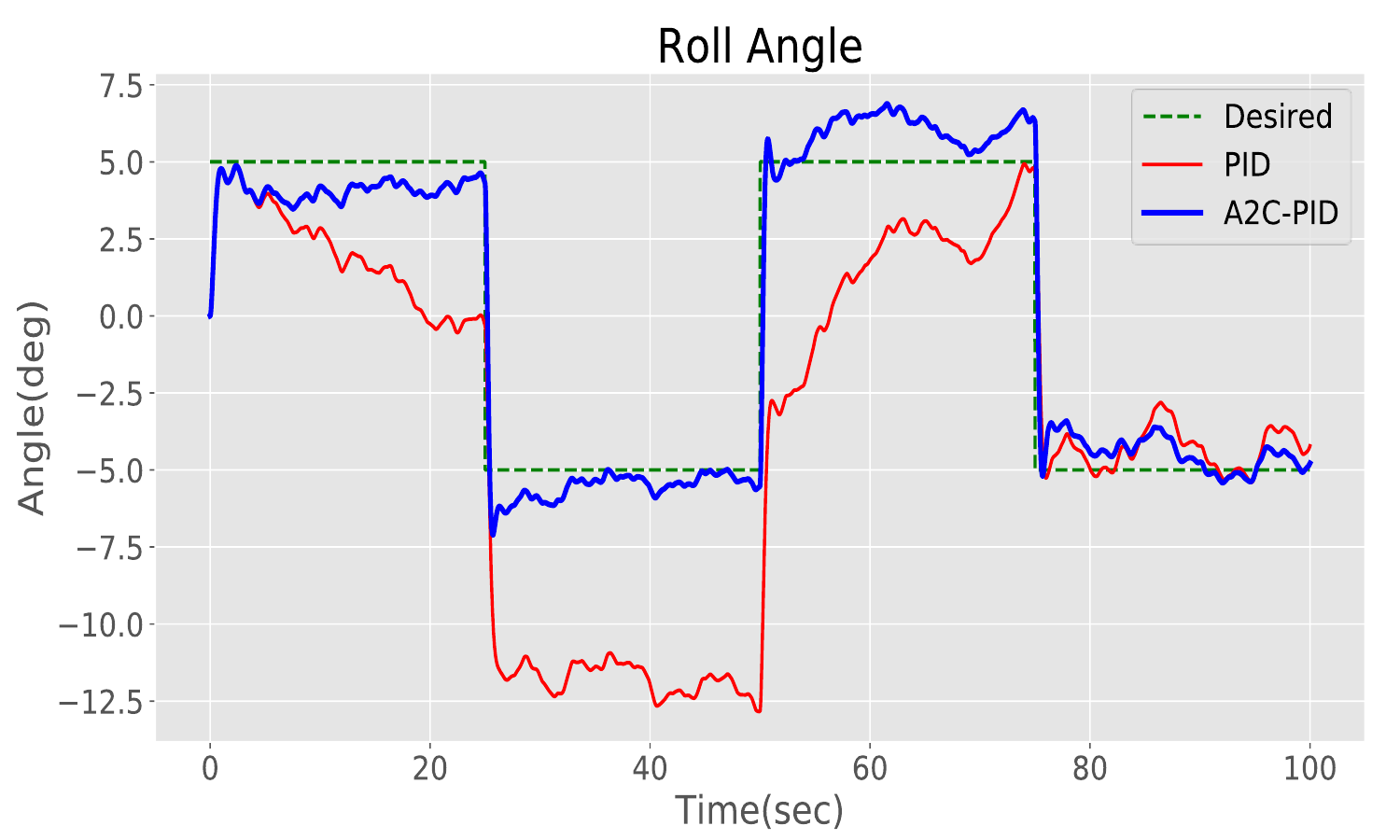}
    \caption{Tracking of the Roll ($\phi$) angle using the proposed method and the PID controller.}
    \label{roll}
\end{subfigure}

\vspace{0.3cm}

\begin{subfigure}{0.48\linewidth}
    \centering
    \includegraphics[width=\linewidth]{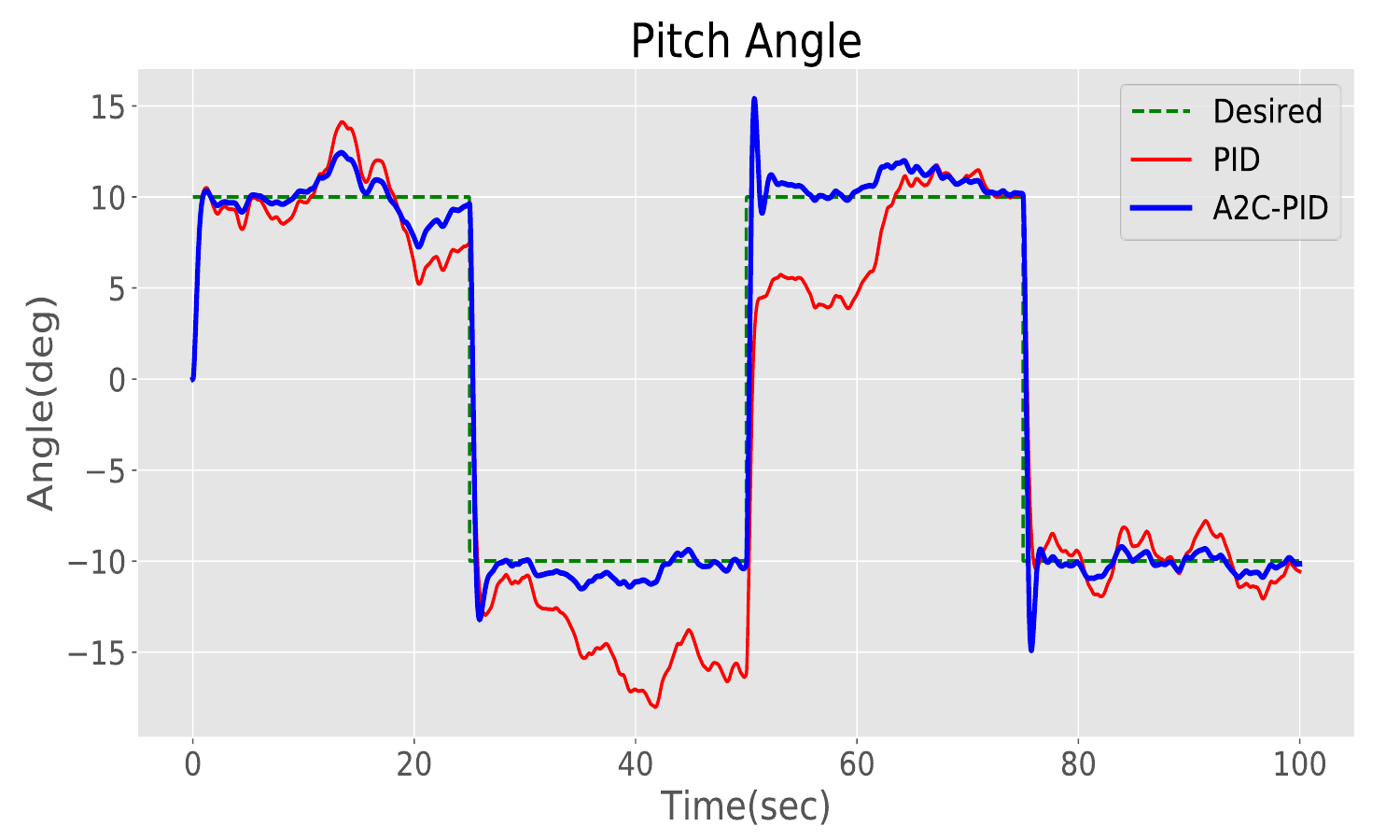}
    \caption{Tracking of the Pitch ($\theta$) angle using the proposed method and the PID controller.}
    \label{pitch}
\end{subfigure}
\hfill
\begin{subfigure}{0.48\linewidth}
    \centering
    \includegraphics[width=\linewidth]{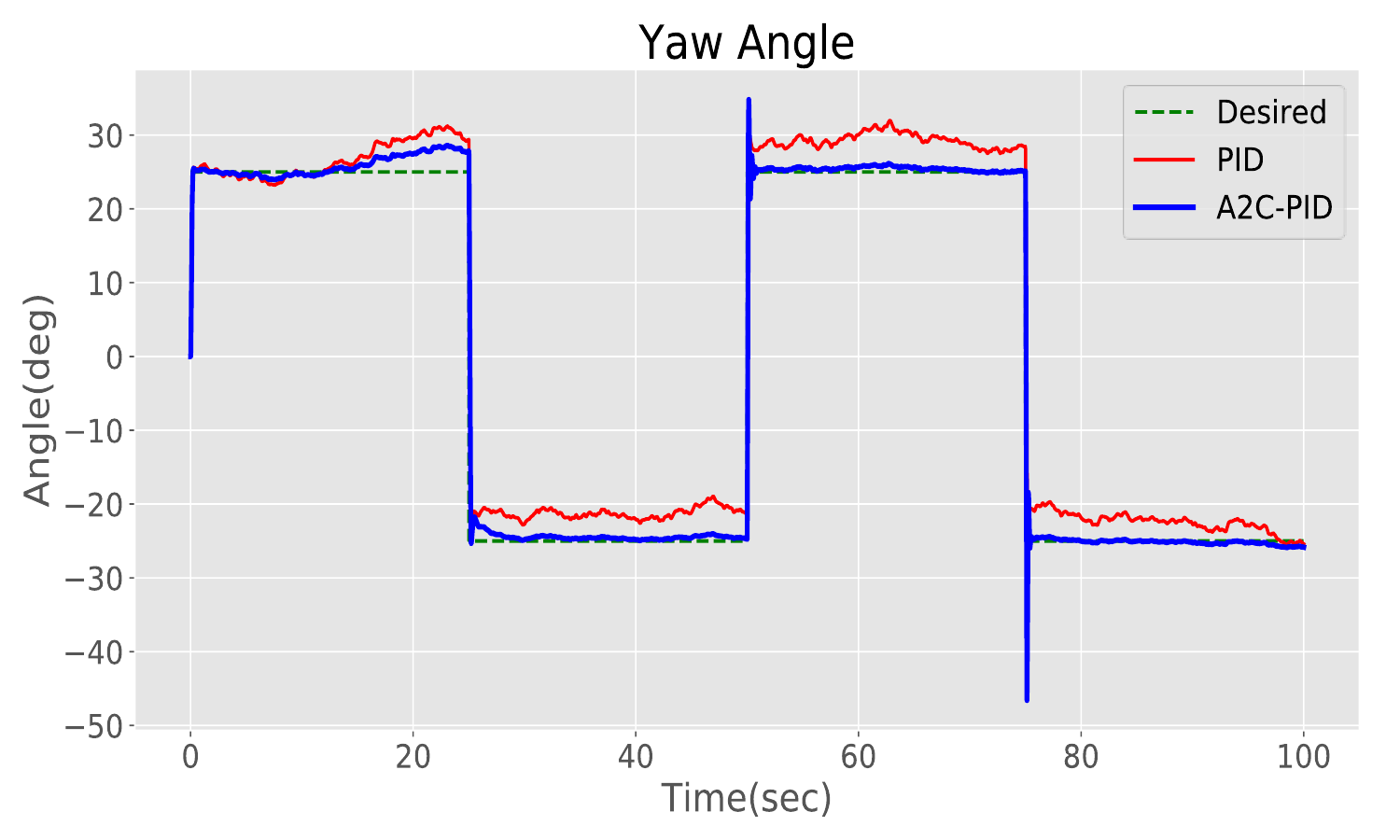}
    \caption{Comparison of Yaw ($\psi$) angle tracking between the proposed method and the conventional PID controller.}
    \label{yaw}
\end{subfigure}

\caption{Comparison of the proposed controller and conventional PID under disturbance conditions.}
\label{disturbance_results}
\end{figure*}

To further challenge the performance of the proposed method, we investigated the impact of wind gust disturbances on the quadcopter’s attitude. Gaussian–Markov equations~\cite{ding2021} were adopted, as given by:
\begin{equation}
	\dot{d} = -\frac{1}{\tau_s} d + \rho B_w q_w,
	\label{dist_eq}
\end{equation}
where $q_w$ is an independent constant with zero mean, $\tau_s = 0.3$ is the correlation time of the wind, $B_w$ is the turbulence input identity matrix, and $\rho = 0.5$ is the scalar weighting factor. Equation~\ref{dist_eq} is commonly referred to as a \emph{shaping filter} for wind gusts. It was solved using the ODE45 method in conjunction with Eq.~\ref{quad eqs}, and the resulting disturbances are shown in Fig.~\ref{dist}. The disturbances, initialized at zero, are sufficiently large to affect the performance of conventional controllers.  

The proposed controller demonstrates robustness against disturbances across all attitude angles, outperforming the conventional PID controller. After some time, the PID controller is unable to compensate for the increasing error because its integrator gain is constant. In contrast, the proposed method adapts its gains dynamically, thereby mitigating error growth. In practice, the controller learns to adjust gains using the error, error rate, control input, and past system states.  

To quantitatively compare the performance of the proposed method with that of the conventional PID controller, the Root Mean Square Error (RMSE) of each attitude angle was computed for both controllers. The results are summarized in Table~\ref{rmse_table}.  Accordingly, A2CPID method could significantly reduce the RMSE for all attitude angles, compared to the PID controller.

\begin{table}[h]
\centering
\caption{Comparison of RMSE values for PID and proposed A2CPID controllers.}
\label{rmse_table}
\begin{tabular}{ccc}
	\hline
	\textbf{Attitude RMSE} (deg) & \textbf{PID} & \textbf{A2CPID} \\
	\hline
	RMSE($\phi$)   & 4.49 & 1.20 \\
        RMSE($\theta$) & 3.85 & 1.83 \\
	RMSE($\psi$)   & 2.50 & 0.35 \\
	\hline
\end{tabular}
\end{table}

\section{Conclusion}
This research presented a novel self-tuning PID control approach based on a hybrid neural architecture employing the actor–critic method. The proposed controller adaptively tunes PID gains and performs state identification in an online manner. Owing to its straightforward structure, the method is applicable to real SISO systems while maintaining scalability to more complex scenarios. The framework leverages neural networks in combination with the Adam optimizer, ensuring fast and reliable training. Simulation results demonstrated that the controller can effectively track complex trajectories, even with randomly initialized weights. Moreover, it exhibited robustness to mass uncertainty and wind gust disturbances, maintaining accurate altitude and attitude control. Overall, the proposed method significantly outperforms conventional PID controllers with fixed gains, highlighting its potential for deployment in real-world quadcopter applications.

\section{Acknowledgment}
This research was conducted as part of the author’s master’s thesis and without any sponsorship.

\bibliographystyle{ieeetr}
\bibliography{references}

\end{document}